\documentclass[twocolumn]{aa}  

\newcommand\dustpy{\texttt{DustPy}}
\newcommand{\revision}[1]{#1}

\usepackage{xcolor}

\usepackage[T1]{fontenc}
\usepackage[english]{babel}

\usepackage[fleqn]{amsmath}
\usepackage{amssymb}
\usepackage{bbold}

\usepackage{savesym}
\savesymbol{tablenum}
\usepackage{siunitx}
\restoresymbol{SIX}{tablenum}

\usepackage{xcolor}
\definecolor{xlinkcolor}{cmyk}{1,1,0,0}
\PassOptionsToPackage{hyphens}{url}
\usepackage[
    bookmarks=true,
    pdfnewwindow=true,
    colorlinks=true,
    linkcolor=xlinkcolor,
    citecolor=xlinkcolor,
    filecolor=xlinkcolor,
    urlcolor=xlinkcolor,
    final=true,
]{hyperref}

\makeatletter
\renewcommand*\aa@pageof{, page \thepage{} of \pageref*{LastPage}}
\makeatother

\usepackage{graphicx}
\usepackage{txfonts}
\begin{document}

\title{Leaky dust traps: How fragmentation impacts dust filtering by planets}

\author{
    Sebastian Markus Stammler\inst{1}
    \and
    Tim Lichtenberg\inst{2}
    \and
    Joanna Dr{\k{a}}{\.z}kowska\inst{3}
    \and
    Tilman Birnstiel\inst{1, 4}
}

\institute{
    University Observatory, Faculty of Physics, Ludwig-Maximilians-Universität München, Scheinerstr. 1, 81679, Munich, Germany
    \and
    Kapteyn Astronomical Institute, University of Groningen, P.O. Box 800, 9700 AV Groningen, The Netherlands
    \and
    Max Planck Institute for Solar System Research, Justus-von-Liebig-Weg 3, 37077 Göttingen, Germany
    \and
    Exzellenzcluster ORIGINS, Boltzmannstr. 2, D-85748 Garching, Germany
 }

\abstract{
The nucleosynthetic isotope dichotomy between carbonaceous (CC) and non-carbonaceous (NC) meteorites has been interpreted as evidence for spatial separation and the coexistence of two distinct planet-forming reservoirs for several million years in the solar protoplanetary disk. The rapid formation of Jupiter’s core within one million years after the formation of calcium-aluminium-rich inclusions (CAIs) has been suggested as a potential mechanism for spatial and temporal separation. In this scenario, Jupiter's core would open a gap in the disk and trap inward-drifting dust grains in the pressure bump at the outer edge of the gap, separating the inner and outer disk materials from each other. We performed simulations of dust particles in a protoplanetary disk with a gap opened by an early-formed Jupiter core, including dust growth and fragmentation as well as dust transport, using the dust evolution software \dustpy{}. Our numerical experiments indicate that particles trapped in the outer edge of the gap rapidly fragment and are transported through the gap, contaminating the inner disk with outer disk material on a timescale that is inconsistent with the meteoritic record. This suggests that other processes must have initiated or at least contributed to the isotopic separation between the inner and outer Solar System.
}


\maketitle

\section{Introduction} \label{sec:intro}

Recent high-precision isotopic measurements reveal a dichotomy between carbonaceous (CC) and non-carbonaceous (NC) meteorites indicating that they were formed in separate reservoirs within the early Solar System \citep{2007ApJ...655.1179T,2009Sci...324..374T,2009ApJ...702.1118L,warren2011E&PSL.311...93W,2020SSRv..216...27M,2020SSRv..216...55K}. \citet{kruijer2017PNAS..114.6712K} and \citet{2018ApJS..238...11D} argue that these reservoirs must have been well separated for at least two million years without interchanging solid material, and propose the rapid formation of Jupiter's core opening a gap in the protoplanetary disk as the possible mechanism that prevents the mixing of the two reservoirs. The physical origin of the isotopic separation is a potential critical clue to the timescales of planet formation in both the inner and outer Solar System \citep{2018SSRv..214..101N}, and thus ultimately the origin of the chemical abundances in the terrestrial planets and similar exoplanets \citep{2022arXiv220310056K,2022arXiv220310023L}.

This concept of a gap-opening Jupiter preventing dust reservoir mixing, however, strongly depends on the evolution of the dust flux during the evolution of the protoplanetary disk. Dust particles in protoplanetary disks are subject to gas drag and drift \citep{whipple1973NASSP.319..355W, weidenschilling1977MNRAS.180...57W, takeuchi2002ApJ...581.1344T}. The radial dust velocity is given by
\begin{equation}
    v_\mathrm{d} = v_\mathrm{g}\frac{1}{\mathrm{St}^2+1} + 2v_\mathrm{P}\frac{\mathrm{St}}{\mathrm{St}^2+1}.
    \label{eqn:vdust}
\end{equation}
The Stokes number $\mathrm{St}$ is an aerodynamic measure and proportional to the particle size. Small particles with small Stokes numbers are dragged along with the gas with velocity $v_\mathrm{g}$, as can be seen by \autoref{eqn:vdust}. The gas, in contrast to the dust, is pressure supported and orbits the star at sub-Keplerian velocities in a typical smooth disk with inward pointing pressure gradient. The dust particles, on the other hand, are not pressure supported, exchange angular momentum with the gas, and drift in the direction of the pressure gradients. Intermediate particle sizes are most affected by this effect. Small particles are tightly coupled to the gas, while large particles are completely decoupled. From \autoref{eqn:vdust} it can be seen that a particle with a Stokes number of unity will experience maximum drift in the direction of the pressure gradient with velocity $v_\mathrm{P}$, which is given by
\begin{equation}
    v_\mathrm{P} = \frac{1}{2} \frac{c_\mathrm{s}^2}{v_\mathrm{K}} \frac{\partial \log P}{\partial \log r},
\end{equation}
with the sound speed $c_\mathrm{s}$, pressure $P$, and the Keplerian velocity $v_\mathrm{K}$. Particles typically grow to maximum sizes with Stokes numbers between $10^{-2}$ to $10^{-1}$ \citep[see][]{birnstiel2012A&A...539A.148B}, depending on the disk parameters, and are therefore affected by radial drift.

Growing planets can perturb the pressure structure in the disk by opening a gap in the gas \citep{paardekooper2006A&A...453.1129P, rice2006MNRAS.373.1619R}. At the outer edge of the gap the pressure gradient reverses and  points outward. If the pressure pertubation is large enough, large dust pebbles that are affected by drift can be prevented from crossing the gap. The planetary mass at which the pressure pertubation is large enough to stop particle drift is called \emph{pebble isolation mass} \citep[see][]{lambrechts2014A&A...572A..35L, bitsch2018A&A...612A..30B} and is given by \citet{drazkowska2022arXiv220309759D} as
\begin{equation}
    M_\mathrm{iso} \simeq  25\,M_\oplus \left(\frac{H_\mathrm{P}/r}{0.05}\right)^3 \frac{M_\star}{M_\odot}.
\end{equation}

NASA's Juno mission estimated that Jupiter's core  has a mass of up to $25\,M_\oplus$ \citep{wahl2017GeoRL..44.4649W}; it  would  therefore have been able to open a gap and stop the flux of dust pebbles in the disk. A rapid formation of Jupiter's core could therefore explain two isolated dust reservoirs with the dust in the outer disk forming the carbonaceous and the dust in the inner disk the non-carbonaceous bodies in the Solar System.

\citet{drazkowska2019ApJ...885...91D}, however, showed in two-dimensional hydrodynamic simulations of gas and dust including collisional dust evolution that the pressure bump at the outer edge of planetary gaps   shows an accumulation of large dust pebbles, and also of small dust particles. However, in contrast to the large pebbles, these small particles are not trapped by the pressure bump; they are produced in situ by collisions of large particles leading to fragmentation. These small fragments can escape the pressure bump due to diffusion and gas drag. The equations of motion of the dust particles are given by
\begin{equation}
    \frac{\partial}{\partial t} \Sigma_\mathrm{d} + \frac{1}{r} \frac{\partial}{\partial r} \left[ r \Sigma_\mathrm{d} v_\mathrm{d} - r D \Sigma_\mathrm{g} \frac{\partial}{\partial r} \left( \frac{\Sigma_\mathrm{d}}{\Sigma_\mathrm{g}} \right) \right] = 0,
    \label{eqn:dust_transport}
\end{equation}
with the dust diffusivity given by \citet{youdin2007Icar..192..588Y} as
\begin{equation}
    D = \frac{\delta_\mathrm{r} c_\mathrm{s}^2}{\Omega_\mathrm{K}} \frac{1}{\mathrm{St}^2+1},
    \label{eqn:dust_diff}
\end{equation}
where $\delta_\mathrm{r}$ is a free parameter that defines the strength of radial dust diffusion. Small particles are therefore most affected by diffusion. If the diffusivity is high enough, these small particles can diffuse out of the pressure maximum and are dragged with the gas through the gap. If this is the case, the inner disk would be contaminated with dust from the outer disk, negating the idea of two distinct dust reservoirs separated by an early-formed Jupiter core.

In this letter we test this hypothesis. In \autoref{sec:toy_model} we present a toy model that initially has dust placed only outside of the planet to show, as a proof of concept, that solid material can penetrate planetary gaps if the dust is subject to fragmentation and diffusion. In \autoref{sec:full} we investigate the influence of the planetary mass and the dust diffusivity on the dust permeability of planetary gaps. In \autoref{sec:time_evo} we present models with a realistic evolution of the planetary mass, as  has been suggested for Jupiter, for models with both fragmentation and bouncing. Finally, in \autoref{sec:discussion} we discuss our results, and   in \autoref{sec:conclusion} we present our conclusions.

\section{Toy model} \label{sec:toy_model}

\begin{figure*}[tb]
    \centering
    \includegraphics[width=\linewidth]{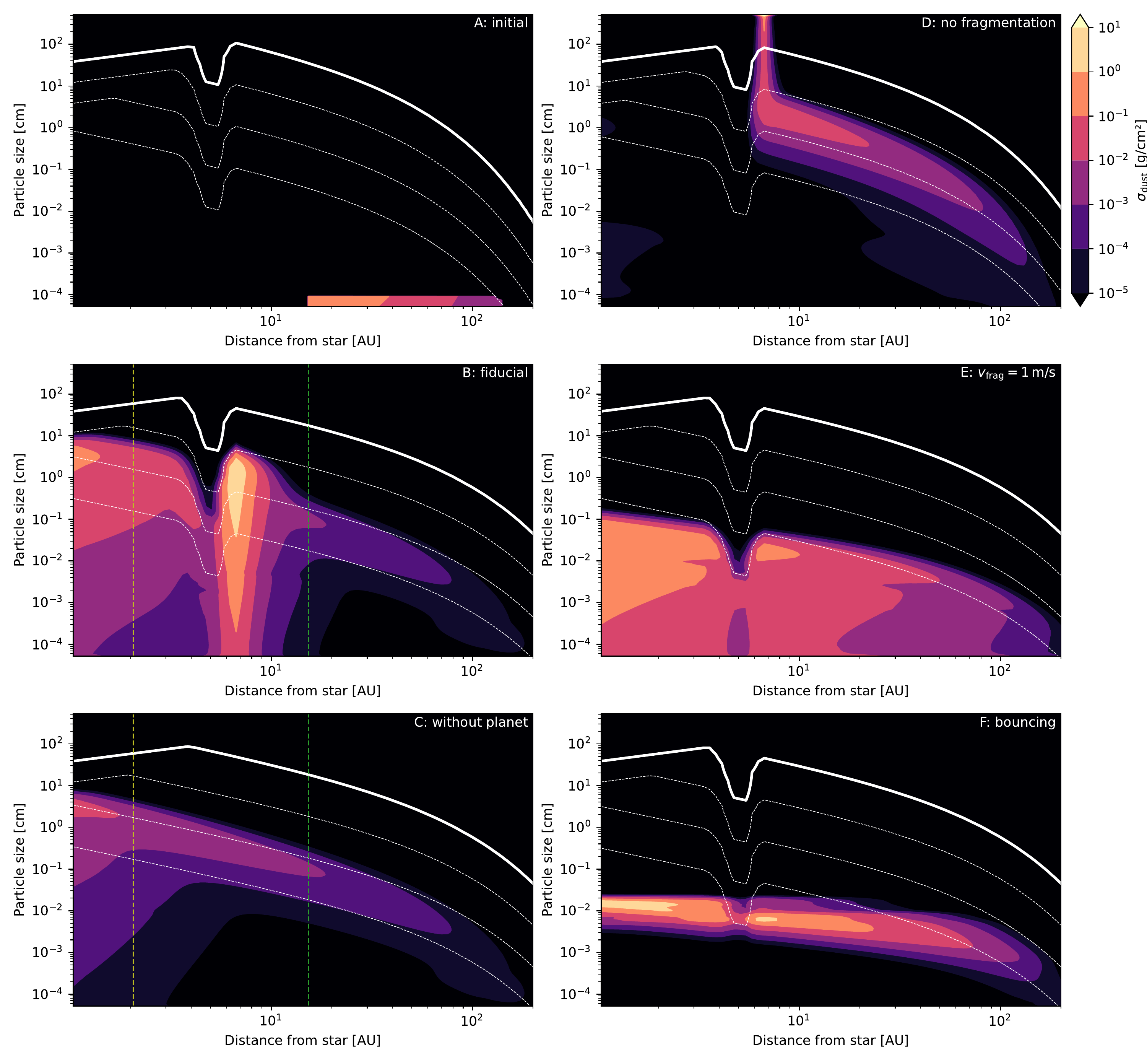}
    \caption{Comparison of different toy models with dust initially only present outside of \SI{15}{AU}. Panel A: Initial dust distribution. The white lines correspond to Stokes numbers of $\mathrm{St} = \left\{ 10^{-3}, 10^{-2}, 10^{-1}, 10^0\right\}$;  the bold white line corresponds to $\mathrm{St}=1$. All other panels show snapshots of models at \SI{1}{Myr}. Panel B: Fiducial toy model with a Saturn-mass planet at \SI{5}{AU} and a fragmentation velocity of \SI{10}{m/s}. Panel C: Model without a planet. The vertical dashed lines are the location at which the dust flux is measured in \autoref{fig:flux_toymodel}. Panel D: Model without fragmentation. Panel E: Model with a reduced fragmentation velocity of \SI{1}{m/s}. Bottom right: Model with bouncing instead of fragmentation.}
    \label{fig:results_toymodel}
\end{figure*}

To investigate the influence of a planet on the dust flux in the inner disk, we modeled dust coagulation and transport in a protoplanetary disk with a planet opening a gap at $5\,\mathrm{AU}$ using the dust evolution software \texttt{DustPy}\footnote{\dustpy{} v1.0.1 was used for the simulations presented in this work.} \citep{stammler2022ApJ...935...35S}. In a first simplified toy model, we initialized the disk only with dust outside of a Saturn-mass planet. Therefore, any dust flux measured inside the planet must have crossed the gap. We used this simplified model to investigate different scenarios: dust growth limited by fragmentation, dust growth limited by bouncing, and unlimited dust growth. Furthermore, we compare the toy model to a model without a gap.

We initialized the gas surface density with the self-similar solution of \citet{lynden-bell1974MNRAS.168..603L}
\begin{equation}
    \Sigma_\mathrm{g} \left( r \right) = \frac{M_\mathrm{disk}}{2\pi r_\mathrm{c}^2} \left( \frac{r}{r_\mathrm{c}} \right)^{-1} \exp \left[ - \frac{r}{r_\mathrm{c}} \right]
\end{equation}
with a cutoff radius of $r_\mathrm{c} = 30\,\mathrm{AU}$ and an initial disk mass of $M_\mathrm{disk} = 0.05\,M_\odot$. We imposed a gap onto this gas surface density profile originating from a Saturn-mass planet located at $5\,\mathrm{AU}$, for which we used the gap profile fits provided by \citet{kanagawa2017PASJ...69...97K}. To maintain this gap profile $F\left(r\right)$ throughout the simulation we imposed the inverse of this profile onto the turbulent viscosity parameter $\alpha$ since the product of gas surface density and viscosity is constant in quasi steady state:
\begin{equation}
    \label{eqn:alpha}
    \alpha \left( r \right) = \frac{\alpha_0}{F\left(r\right)}.
\end{equation}
In the default setup, we use $\alpha_0=\delta_\mathrm{r}=10^{-3}$. We note that this change in $\alpha\left(r\right)$ does not affect the turbulent diffusion of the dust particles since $\delta_r$ is a constant in our models.

We initialized the dust surface density with a constant gas-to-dust ratio of $100$ and the dust size distribution according \citet{mathis1977ApJ...217..425M} as $n\left( a \right) = a^{-3.5}$ with a maximum initial particle size of $1\,\mu\mathrm{m}$. In the toy model we initially had dust only outside of $15\,\mathrm{AU}$. \texttt{DustPy} simulates dust growth by solving the Smoluchowski equation of a dust mass distribution. Dust transport is simulated by solving \autoref{eqn:dust_transport} for every dust size individually.

The gas surface density is evolved by solving the viscous advection-diffusion equation
\begin{equation}
    \frac{\partial}{\partial t} \Sigma_\mathrm{g}  + \frac{1}{r} \frac{\partial}{\partial r} \left( r\Sigma_\mathrm{g}v_\mathrm{g} \right) = 0
,\end{equation}
with the gas velocity given by \citet{lynden-bell1974MNRAS.168..603L} as
\begin{equation}
    v_\mathrm{g} = - \frac{3}{\Sigma_\mathrm{g}\sqrt{r}} \frac{\partial}{\partial r} \left( \Sigma_\mathrm{g} \nu \sqrt{r} \right)
\end{equation}
and the kinematic viscosity given by
\begin{equation}
    \nu = \alpha c_\mathrm{s} H_\mathrm{P}
\end{equation}
with the sound speed $c_\mathrm{s} = \sqrt{k_\mathrm{B}T/\mu}$, the pressure scale height $H_\mathrm{P} = c_\mathrm{s}/\Omega_\mathrm{K}$, and the viscosity parameter $\alpha$ given by \autoref{eqn:alpha}.

We  ran five different flavors of the toy model: one with a fragmentation velocity of $v_\mathrm{frag} = 10\,\mathrm{m/s}$ (fiducial), one with no fragmentation at all, one with a fragmentation velocity of $1\,\mathrm{m/s}$, one with bouncing as described by \citet{windmark2012A&A...540A..73W}, and one without a gap (i.e., $F\left(r\right)=1)$. \revision{In the default collision model used by \dustpy{} particles fragment once their relative collision velocities exceed the fragmentation velocity. Fragmenting collisions of equal size particles lead to catastrophic fragmentation of both collision partners. If the target particle is significantly larger, only the projectile particle fragments entirely while eroding mass off the target particle \citep{schraepler2018ApJ...853...74S, hasegawa2021ApJ...915...22H}. The transition between pure sticking and fragmentation is smooth, since \dustpy{}  assumes a velocity distribution of possible collision velocities. For details on the collision model we refer to \citet{stammler2022ApJ...935...35S}.}

Panel A of \autoref{fig:results_toymodel} shows the initial dust distribution with dust only located outside of $15\,\mathrm{AU}$ with particles sizes up to $1\,\mu\mathrm{m}$. Panel B shows the fiducial simulation with a Saturn-mass planet at $5\,\mathrm{AU}$ and the fragmentation velocity $v_\mathrm{frag} = 10\,\mathrm{m/s}$ after $1\,\mathrm{Myr}$. Particles trapped in the pressure bump outside the planetary gap can reach sizes with Stokes numbers of up to $\mathrm{St}=10^{-1}$ corresponding to particle sizes of a few centimeters. It can be seen that even small particles are accumulated in the pressure bump, even though their Stokes numbers are too small to be affected by drift. These small dust particles are produced by collisional fragmentation of larger particles trapped in the bump. They diffuse out of the bump and are dragged with the gas contaminating the inner disk with outer disk material. It can be seen that particles with Stokes numbers of about $\mathrm{St}=10^{-2}$, corresponding to particle sizes of a few millimeters, can diffuse through the gap into the inner disk. \revision{Particles in the inner disk can again grow to centimeter sizes and can contribute to phenomena like the streaming instability or pebble accretion.} Panel C shows a simulation with identical initial conditions, but without a planet opening a gap.

\begin{figure}[tb]
    \centering
    \includegraphics[width=\linewidth]{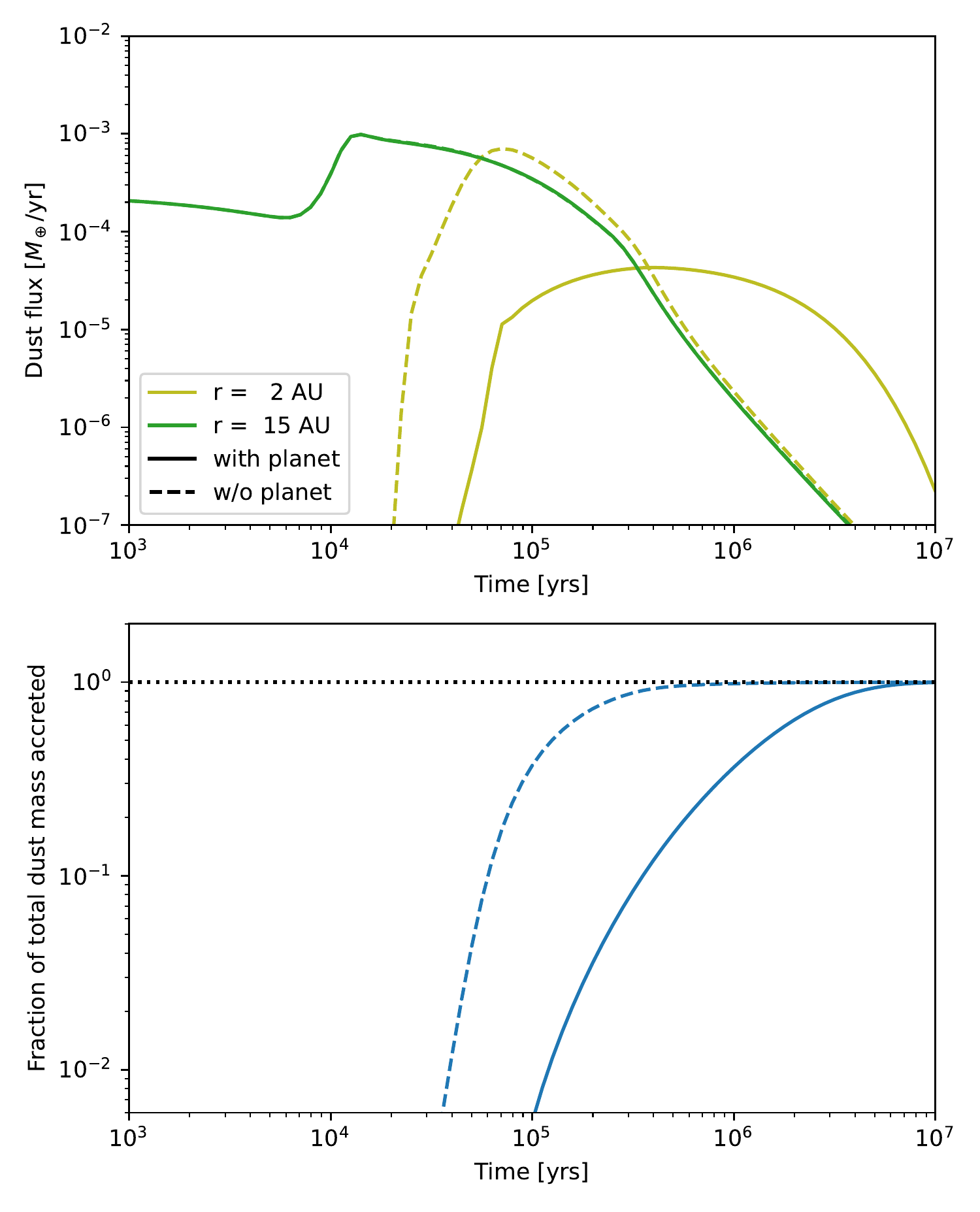}
    \caption{Comparison of dust fluxes and the fraction of total accreted dust mass in the toy model with and without planet. Top: Comparison of the dust flux in the inner disk (at $\SI{2}{AU}$) and outer disk (at $\SI{15}{AU}$) in the toy model with a Saturn-mass  planet at $5\,\mathrm{AU}$ (panel B in \autoref{fig:results_toymodel}) and a model without a planet (panel C in \autoref{fig:results_toymodel}). The two green 15\,AU lines overlap. Bottom: Total dust mass accreted through the inner disk over time in the model with (solid line) and without (dashed line) a planet.}
    \label{fig:flux_toymodel}
\end{figure}

The dust fluxes measured in the outer disk at \SI{15}{AU} are identical in both simulations with the solid and dashed green lines overlapping in \autoref{fig:flux_toymodel}. The fluxes in the inner disk at \SI{2}{AU}, however, differ in the two simulations. The onset of dust flux in the inner disk in the simulation with a planet is delayed by about $20\,000\,\mathrm{yr}$ compared to the simulations without a planet. Without a planet, the large dust particles can freely drift into the inner disk. With a planet, however, they are first trapped in the pressure bump at the outer edge of the gap, fragment down to smaller sizes, and diffuse out of the pressure bump before the gas can drag them into the inner disk where they grow to larger particles again. Due to this delayed processing, the maximum dust flux is reduced by about one order of magnitude. The duration, however, is prolonged such that the total mass of dust flowing through the inner disk is identical after $10\,\mathrm{Myr}$, as can be seen in the bottom panel of \autoref{fig:flux_toymodel}. The Saturn-mass planet did not separate the inner from the outer disk material, but only delayed the material transport.

Panel D of \autoref{fig:results_toymodel} shows a simulation without fragmentation. In this scenario, particles sizes are limited only by the radial drift, \revision{consistent with the model presented by \citet{kobayashi2021ApJ...922...16K}}. In the center of the pressure bump, the pressure gradient is zero and the growth is, in principle, unlimited until the particles accumulate at the upper end of the simulation grid. This scenario most closely represents the separation of inner and outer dust reservoirs with  very few particles being able to diffuse through the gap because they were not able to grow to large particles quickly enough. It is, however, rather unlikely that the dust particles do not fragment or get eroded at some point given the relative velocities they typically experience \citep[see][]{blum1993Icar..106..151B, wada2009ApJ...702.1490W, schraepler2018ApJ...853...74S}.

Panel E of \autoref{fig:results_toymodel} shows a model with a fragmentation velocity of $1\,\mathrm{m/s}$  as indicated by recent experiments  \citep[see, e.g.,][]{blum2018SSRv..214...52B, gundlach2018MNRAS.479.1273G, musiolik2019ApJ...873...58M}. In this case, the particles cannot reach particles sizes large enough to be efficiently trapped in the pressure bump.

The objective is therefore to halt particle growth without producing small particles. This can be achieved if the growth is limited by bouncing, when particles simply bounce off of each other without growing or fragmenting. Panel F of \autoref{fig:results_toymodel} shows a simulation with the bouncing barrier implemented as  described by \citet{windmark2012A&A...540A..73W}. In this model, bouncing starts when the relative velocity reaches a few centimeters per second. In this case, however, the particles only reach sizes of a few $\SI{100}{\micro m,}$ corresponding to Stokes numbers lower than $10^{-3}$, which is too small to be efficiently trapped in the pressure bump created by the planet. The particles can diffuse through the gap and contaminate the inner disk.

\section{Full disk models} \label{sec:full}

\begin{figure}[tb]
    \centering
    \includegraphics[width=\linewidth]{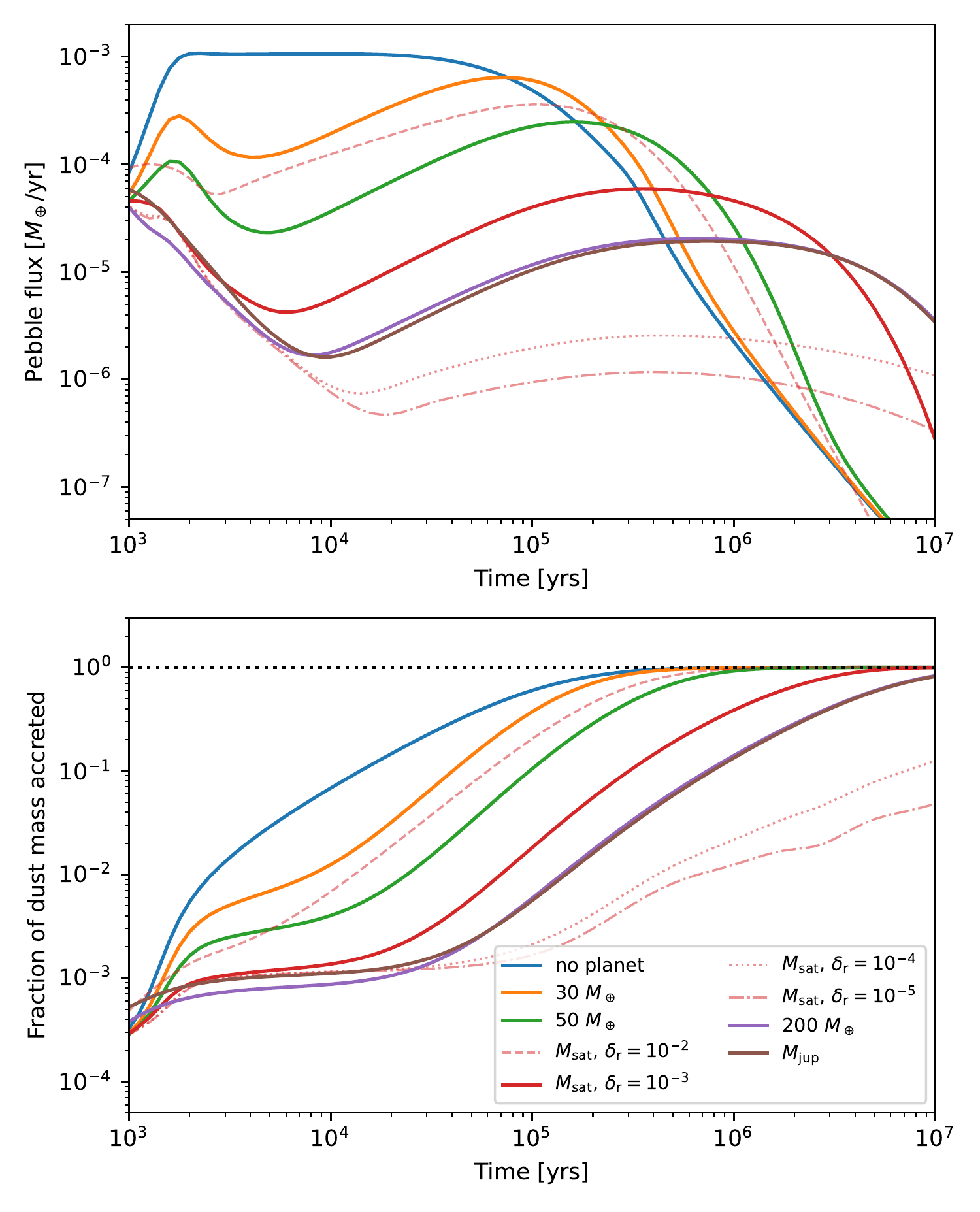}
    \caption{Comparison of dust fluxes and the fraction of total accreted dust mass in the full disk models for various planet masses and dust diffusivities. Top: Dust flux through the planetary gap in models with different planet masses. The blue line is for a model without a planet. The dashed, dotted, and dash-dotted red lines show additional simulations with a Saturn-mass planet for different radial dust diffusivity parameters $\delta_r$. Bottom: Total fraction of outer dust mass accreted through the planetary gap.}
    \label{fig:results_fullmodel}
\end{figure}

The toy model in \autoref{sec:toy_model} served as a proof of concept that planets do not prevent dust flux if particles are subject to fragmentation. In this section we discuss full disk models with different planet masses where  dust is initialized in the entire disk to investigate the dust permeability of the gap. The top panel of \autoref{fig:results_fullmodel} shows the dust flux through the planetary gap for different planet masses, from 30 Earth masses to one Jupiter mass. In the case of a Saturn-mass planet, we additionally performed simulations with different dust diffusivity parameters $\delta_r$ (see \autoref{eqn:dust_diff}). The bottom panel of \autoref{fig:results_fullmodel} shows the total fraction of outer disk dust material that has been accreted through the gap over time. In all the models the planets have their respective masses already from the beginning of the simulations.

The lowest planetary mass considered here is $30\,M_\oplus$, which is already higher than the upper estimate of Jupiter's core mass. The highest mass considered is $1\,M_\mathrm{jup}$. The  lowest planetary mass is not capable of efficiently suppressing the dust flux through the gap. After about $300\,000\,\mathrm{yr}$ almost the entire dust mass (horizontal line in bottom panel) of the outer disk has been accreted through the gap. Increasing the planetary mass simply delays the accretion time, but is not able to prevent accretion. The maximum delay of accretion seems to be achieved already with a $200\,M_\oplus$ planet. Increasing the planet mass further to a Jupiter-mass  planet does not significantly change the accretion history. At the end of the simulation at $10\,\mathrm{Myr}$ about $80\,\%$ of the dust mass has been accreted through the gap.

The dust diffusivity $\delta_r$ has a more significant influence on the accretion. Increasing the diffusivity by a factor of $10$ to $\delta_r=10^{-2}$ in the Saturn-mass simulation has the same effect as reducing the planet mass by a factor of about $2$, mimicking the accretion history of a $\SI{40}{M_\oplus}$ planet with diffusivity of $\delta_r=10^{-3}$. We note that we only changed $\delta_r$ and kept $\alpha_0=10^{-3}$, and therefore keeping the shape of planetary gap. The relative collision velocities of the dust particles are not affected by this change in $\delta_r$. Decreasing $\delta_r$ by a factor of $10$ is more efficient in retaining the dust than having a Jupiter-mass planet with the default diffusivity. In this case only about $10\,\%$ of the dust mass has been accreted at the end of the simulation after $10\,\mathrm{Myr}$. Lowering the diffusivity even further to $\delta_r=10^{-5}$ reduces the dust permeability further to  about $5\,\%$ of the outer disk mass after $10\,\mathrm{Myr}$. It should be noted,  however,  that the fraction of outer disk material present in the inner disk is usually significantly larger since the inner disk material is accreted onto the star on short timescales and is only re-supplied with outer disk material.

\section{Time-dependent planet mass} \label{sec:time_evo}

\begin{figure*}[tb]
    \centering
    \includegraphics[width=\linewidth]{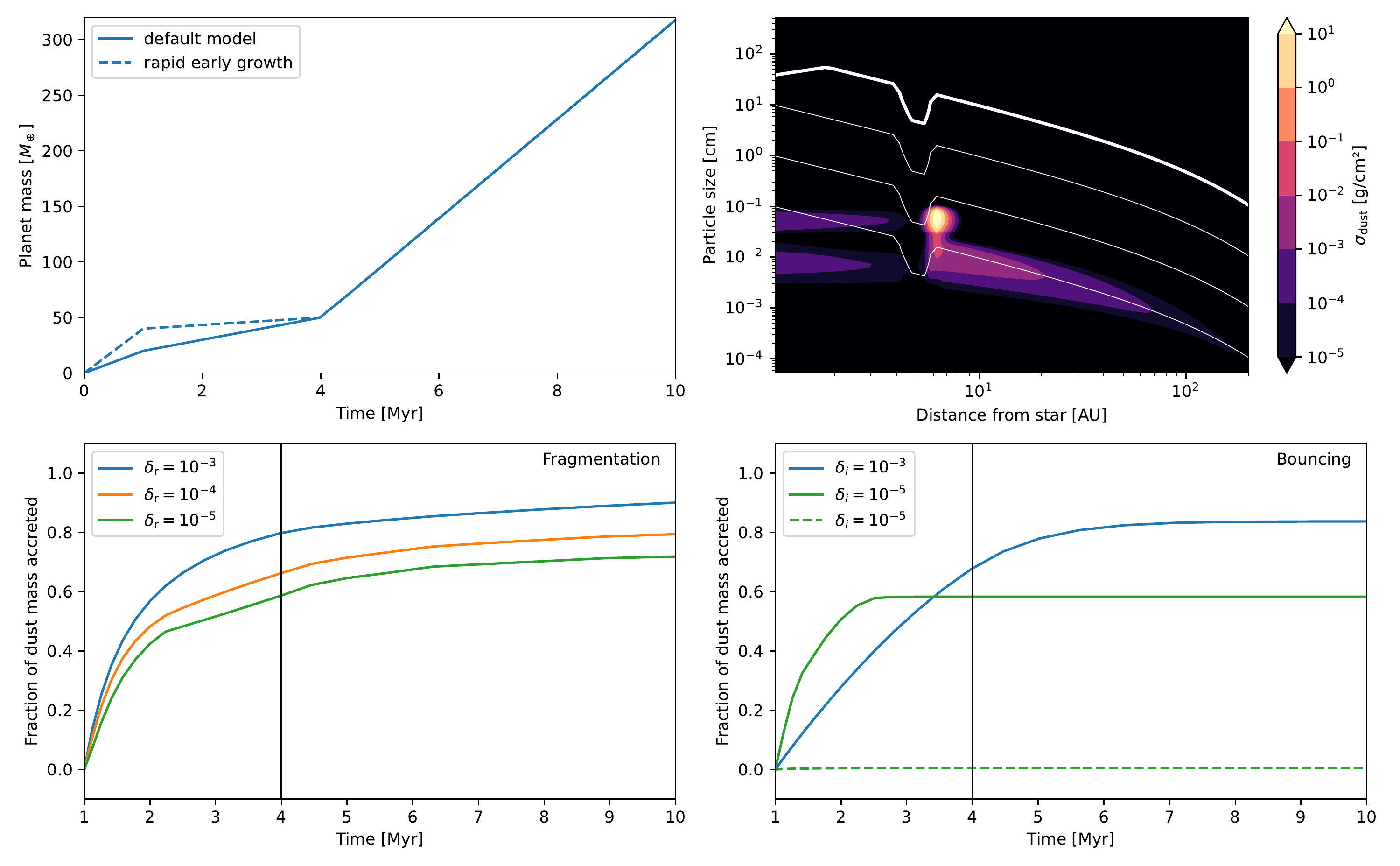}
    \caption{Comparison of models with fragmentation and bouncing with variable planet masses. \textbf{Top left:} Evolution of the planetary mass in the time-dependent model. The solid line shows the default model where the planet reaches $\SI{20}{M_\oplus}$ at $\SI{1}{Myrs}$. The dashed line shows the evolution in a model with rapid early growth in which the planet reaches $\SI{40}{M_\oplus}$ at $\SI{1}{Myr}$. \textbf{Bottom left:} Fraction of outer disk dust mass accreted through the gap after $\SI{1}{Myr}$ in the default planetary mass evolution model for different values of dust diffusivity $\delta_r$ with fragmentation limited growth. \textbf{Bottom right:} Solid lines show the fraction of outer disk material accreted through the gap after $\SI{1}{Myr}$ for bouncing limited growth for different values of the $\delta_i$ parameters in the default planetary growth model. The dashed green line shows a model of bouncing limited growth with $\delta_i=10^{-5}$ and rapid early growth of the planet (dashed line in top left panel). The vertical lines indicate $\SI{4}{Myr}$ until which the two reservoirs needed to be separated. \textbf{Top right:} Snapshot of the dust distribution at $\SI{4}{Myr}$ for the model with bouncing limited growth and $\delta_i=10^{-5}$ (dashed green line in bottom right panel). The inner disk is depleted in dust and only supplied with small amounts of outer disk material.}
    \label{fig:results_timeevo}
\end{figure*}

In the previous models we assumed that the planets are fully formed from the beginning of the simulation and the planet mass does not evolve over time. \citet{kruijer2017PNAS..114.6712K} argue that the two dust reservoirs have been separated from about $1\,\mathrm{Myr}$ to $3-4\,\mathrm{Myr}$ after the formation of calcium-aluminium-rich inclusions (CAIs). They therefore claim that Jupiter's core must have been massive enough to open a gap at $1\,\mathrm{Myr}$ and must have reached a mass of about $50\,M_\oplus$ after $4\,\mathrm{Myr}$ to be able to scatter planetesimals from the outer disk to the inner disk where they are observed today in the asteroid belt. We therefore performed simulations with a time-dependent planet mass, as shown in the top left panel of \autoref{fig:results_timeevo}. The solid blue line shows an evolutionary track where the planet reaches $30\,M_\oplus$ after $1\,\mathrm{Myr}$, $50\,M_\oplus$ after $4\,\mathrm{Myr}$, and a final mass of $M_\mathrm{jup}$ at the end of the simulation after $10\,\mathrm{Myr}$.

The bottom left panel of \autoref{fig:results_timeevo} shows the fraction of mass accreted through the planetary gap normalized to the dust mass in the outer disk at $1\,\mathrm{Myr}$ when the planet was massive enough to open a gap. We performed simulations with different values of the dust diffusivity $\delta_r$ between $10^{-5}$ and $10^{-3}$. In the standard run with $\delta_r=10^{-3}$ about $80\,\%$ of the dust mass was accreted through the gap after $4\,\mathrm{Myr}$ when the assembly of the meteorite parent bodies was   completed. Even in the low diffusivity run with $\delta_r=10^{-5}$ about $60\,\%$ of the mass was   accreted though the gap between $1\,\mathrm{Myr}$ and $4\,\mathrm{Myr}$, strongly contaminating the inner disk with dust from the outer reservoir on a system-wide scale. Lowering the dust diffusivity to very low values did not help in keeping the two reservoirs separated since the planet mass is too low in this scenario.

The bottom right panel of \autoref{fig:results_timeevo} shows a model with bouncing instead of fragmentation. As already shown in \autoref{sec:toy_model}, this is not sufficient to stop dust accretion through the gap. Only after $\SI{7}{Myr}$ when the planet already reached a mass of about $\SI{200}{M_\oplus}$ is the gap  deep enough and accretion  halted. Allowing the planet to reach these masses at earlier times, however,  would not change the dust redistribution since these massive planets are able to scatter planetesimals from the outer disk into the inner disk, which is inconsistent with observations from the meteoritic record at these early times \citep{2022ApJ...936L..24D}.

We ran additional models with $\delta_r = \delta_t = \delta_z = 10^{-5}$. The parameters $\delta_t$ and $\delta_z$ are similar to $\delta_r$ and parameterize the strength of turbulent motion and vertical settling of the particles \citep[see][for details]{stammler2022ApJ...935...35S, pinilla2021A&A...645A..70P}. In that way the relative velocities between the particles are reduced, allowing them to grow to larger sizes before being limited by bouncing. They can therefore be trapped by gaps created by lower mass planets. However, even in that case accretion was only halted after abut $\SI{3}{Myr}$ when the planet reached a mass of about $\SI{40}{M_\oplus}$.

We therefore additionally ran a model where the planet reaches a mass of $\SI{40}{M_\oplus}$ already after $\SI{1}{Myr}$. In this case accretion of dust through the gap was efficiently stopped at $\SI{1}{Myr}$. The top right panel of \autoref{fig:results_timeevo} shows a snapshot of this simulation after $\SI{4}{Myr}$. The inner disk is heavily depleted in dust, all of which has been accreted onto the star. The dust mass in the inner disk at this stage was all supplied from the outer disk. Meteoritic bodies formed in the inner disk would therefore be entirely made out of outer disk material.

\section{Discussion} \label{sec:discussion}

Isotopic measurements of meteoritic material indicate that meteorites must have formed in two dust reservoirs that coexisted spatially separated for several million years. The early formation of Jupiter's core has been proposed as natural explanation for the observed separation. A planet exceeding the pebble isolation mass opens a gap in the gas disk creating a pressure bump at the outer edge of the gap, which can trap large dust particles. Two-dimensional hydrodynamic simulations by \citet{drazkowska2019ApJ...885...91D} including dust coagulation and fragmentation showed an overabundance of small dust particles at the location of the pressure bump, which should be too small to be efficiently trapped. These particles were created via a fragmenting collision of large dust pebbles   trapped in the pressure bump. These small dust fragments can diffuse out of the bump and can be dragged by the gas through the gap.

Our simulations in this work suggest that the collisional fragmentation of dust pebbles in pressure bumps and the subsequent diffusion of small fragments can act as a leak for dust traps. As can be seen in \autoref{fig:results_fullmodel}, gaps opened by planets can only delay but not fully prevent dust accretion if particles are subject to fragmentation. To act as an efficient dust barrier, particles need to grow to large pebbles that can be trapped without producing small particles, as shown in   panel D of \autoref{fig:results_toymodel}.

We investigated different planet masses and showed in \autoref{fig:results_fullmodel} that no planet mass was able to completely isolate the inner disk from the outer dust material on timescales that are relevant for the assumed reservoir separation. Even an initial gap formed by a fully-grown Jupiter-mass planet would leak $20\,\%$ of the outer disk material into the inner disk within $1\,\mathrm{Myr}$. Lower proto-Jupiter masses typically lead to complete homogenization within $\sim10^5$ to at most a few $10^6$ yr. This presents a problem for the suggestion that the age differences in carbonaceus and non-carbonaceous meteorites may be used as a tracer to track the growth timescale of proto-Jupiters within the disk \citep{kruijer2017PNAS..114.6712K,2018NatAs...2..873A}; the initial spatial distribution of nucleosynthetic isotopes at the end of disk infall is degenerate with different Jupiter growth tracks in the Jupiter barrier hypothesis. Only significantly lowering the dust diffusivity to a value of $\delta_r = 10^{-5}$ could decrease the dust permeability such that the inner disk is only contaminated with a small percentage of outer disk material. However, isolating the inner disk from dust flux would quickly drain the inner disk from solids that got accreted onto the star, which was also previously noted by \citet{liu2022SciA....8M3045L}. At later stages the dust in the inner disk then consists, to a large degree, of outer disk material that had  slowly diffused through the gap, which is inconsistent with the meteoritic record.

The situation gets worse when using a more realistic evolution of the planetary mass, assuming Jupiter's core reached a mass of $20\,M_\oplus$ after $1\,\mathrm{Myr}$ and $50\,M_\oplus$ after $4\,\mathrm{Myr}$. These masses are not high enough to isolate the inner disk even in models with very low diffusivity. Even in the most optimistic cases at least $60\,\%$ of the outer disk dust was accreted through the planetary gap after $4\,\mathrm{Myr,}$ as can be seen in \autoref{fig:results_timeevo}. However, increasing the core mass even more and earlier would enable Jupiter to scatter outer disk planetesimals into the inner disk, polluting the inner dust reservoir, which has not been accounted for in this simple model. Only in models with bouncing limited growth without small particles, early planetary growth, and reduced relative particle collision velocities can the inner disk   be efficiently isolated from the outer disk, as seen by \autoref{fig:results_timeevo}. In these cases, however, the inner disk is quickly depleted from dust and only re-supplied by a small amount of outer disk material. Meteoritic bodies formed in the inner disk after this point would therefore consist almost entirely of outer disk material.

\citet{drazkowska2019ApJ...885...91D} noted that the shape of planetary gaps in two-dimensional simulations is not axisymmetric, which is ignored in the simple one-dimensional model in this publication. They further noted, however, that the asymmetry at the planet location would increase the dust flux through the gap. \citet{weber2018ApJ...854..153W} compared one- and two-dimensional simulations of dust transport through planetary gaps, and indeed found that gaps in two-dimensional simulations are more permeable to dust particles. Our one-dimensional simulations, therefore, need to be considered as more restrictive. If it is not possible to separate two reservoirs in one-dimensional models, it is less likely to be possible  in higher dimensions.

\revision{We furthermore assumed a dust fragmentation velocity of $10\,\mathrm{m/s}$, which may be rather high even for icy particles, as indicated by  recent laboratory experiments that  suggest values of $1\,\mathrm{m/s}$ \citep[see][]{blum2018SSRv..214...52B, gundlach2018MNRAS.479.1273G, musiolik2019ApJ...873...58M}. Lowering the fragmentation velocity, however, generally decreases the particle sizes, making them even less likely to be trapped in pressure bumps (see panel E in \autoref{fig:results_toymodel}). Other experiments indicate a significantly higher fragmentation velocity \citep[e.g.,][]{kimura2020MNRAS.496.1667K} than the $\SI{10}{m/s}$ used in this work. The exact value of the fragmentation velocity, however, does not significantly influence the problem of inner disk contamination. Either the fragmentation velocity is exceeded, which will lead to pollution of the inner disk with outer disk material (see panel B of \autoref{fig:results_toymodel}). Or the fragmentation velocity is greater than the maximum collision velocity of dust particles in the disk, in which case the particles will efficiently grow to larger particles, which are   trapped in the outer edge of the disk, and which will quickly deplete the inner disk (see panel D in \autoref{fig:results_toymodel}).} \revision{Similarily, the porosity evolution may have an effect on the collisional physics of dust particles \citep[e.g.,][]{suyama2008ApJ...684.1310S, krijt2015A&A...574A..83K, kobayashi2021ApJ...922...16K}. However, as for the fragmentation velocity, the details of the collision model do not have a strong effect on the outcome of the simulation. Either the particles fragment and the inner disk is polluted with outer disk material, or the particles grow unhindered to large particles that are trapped in the pressure bump, which  quickly depletes the inner disk.}

\revision{We furthermore did not consider the formation of planetesimals in the pressure bump in this work. Previous publications have shown that the conditions in pressure maxima at the outer edges of gaps can facilitate planetesimal formation \citep{stammler2019ApJ...884L...5S, miller2021MNRAS.508.5638M} or even the formation of planets \citep{lau2022A&A...668A.170L, jiang2023MNRAS.518.3877J}. One could conceive that small dust fragments could not penetrate the inner disk because they are quickly converted into planetesimals before they could transverse the gap. This would, however, require a nearly perfect planetesimal formation efficiency to efficiently isolate the two dust reservoirs, which has not been observed in previous simulations. Additionally, in simulations planetesimals formed at gap edges  have been shown to quickly ablate \citep{2021A&A...648A.112E}. Enstatite and ordinary chondrites would thus have to be explained by planetesimal formation where the dust is replenished, for instance by late-stage planetesimal collisions in the NC reservoir \citep{2014ApJ...794...91D,2018Icar..302...27L,2022ApJ...927L..22B}.}

This suggests that it is unlikely that the formation of Jupiter could have solely separated both dust reservoirs in the Solar System if the dust particles were subject to fragmentation. This does not only apply to gaps created by planets, but also to other substructures of non-planetary origin where particles are trapped in pressure maxima, as described in \citet{2020NatAs...4..492B}. Other suggested mechanisms to explain the observations include a temporal change in the isotopic content of inward-streaming dust grains \citep{2018Natur.555..507S}, and the formation of multiple distinct planetesimal populations in the inner and outer disk \citep{2021Sci...371..365L,2022NatAs...6...72M,2022NatAs...6..357I,liu2022SciA....8M3045L}. How these physical mechanisms are connected to the structures and gaps seen in ALMA disks \citep{2022arXiv220309818M} and the underlying mechanisms of protoplanet formation \citep{drazkowska2022arXiv220309759D} and differentiation \citep{2022arXiv220310023L} remains to be explored.

\section{Conclusions} \label{sec:conclusion}

Protoplanet-induced gaps in circumstellar disks are not able to efficiently separate dust in the inner disk from dust in the outer disk on million-year timescales if the particles are subject to fragmentation. Particles limited by bouncing without producing small fragments are usually too small to be trapped by pressure bumps. Only significantly reducing the relative collision velocities allows particles to be efficiently trapped in pressure bumps within $\SI{1}{Myr}$, if the planet grew to $\SI{40}{M_\oplus}$. In this case, however, the inner disk is quickly depleted from dust, making it difficult to form meteoritic bodies in situ. Our simulations suggest that other physical mechanisms must have initiated or at least substantially contributed to the large-scale separation of nucleosynthetic isotopes observed in the planetary materials of the inner and outer Solar System.

\begin{acknowledgements}
This project has received funding from the European Research Council (ERC) under the European Union’s Horizon 2020 research and innovation programme under grant agreement No 714769. This project has received funding by the Deutsche Forschungsgemeinschaft (DFG, German Research Foundation) through grants FOR 2634/1 and 361140270. This research was supported by the Munich Institute for Astro-, Particle and BioPhysics (MIAPbP) which is funded by the Deutsche Forschungsgemeinschaft (DFG, German Research Foundation) under Germany's Excellence Strategy – EXC-2094 – 390783311. JD was funded by the European Union under the European Union’s Horizon Europe Research \& Innovation Programme 101040037 (PLANETOIDS). Views and opinions expressed are however those of the authors only and do not necessarily reflect those of the European Union or the European Research Council. Neither the European Union nor the granting authority can be held responsible for them. TL was supported by grants from the Simons Foundation (SCOL Award No. 611576) and the Branco Weiss Foundation.
\end{acknowledgements}

\bibliographystyle{aa}
\bibliography{bibliography,bib_TL}{}

\end{document}